\documentclass[conference]{IEEEtran}

\usepackage{amsmath,amssymb}
\usepackage{mathtools}
\usepackage{graphicx}
\usepackage{booktabs}
\usepackage{multirow}
\usepackage{cite}
\usepackage{url}
\usepackage[vlined,linesnumbered]{algorithm2e}
\RestyleAlgo{ruled}
\usepackage{hyperref}
\usepackage{xcolor}
\usepackage{caption}
\usepackage{subcaption}
\usepackage{amsthm}
\newtheorem{remark}{Remark}
\usepackage{nicefrac}

\usepackage[capitalize,noabbrev,nameinlink]{cleveref}
\crefformat{equation}{(#2#1#3)}
\crefname{algocf}{algorithm}{algorithms}
\Crefname{algocf}{Algorithm}{Algorithms}
\Crefname{assumption}{Assumption}{Assumptions}
\crefrangeformat{assumption}{Assumptions~#3#1#4-#5#2#6}
\crefname{remark}{remark}{remarks}
\Crefname{remark}{Remark}{Remarks}

\usepackage{mdframed}
\newcommand{\contentbox}[2]{
  \begin{mdframed}[backgroundcolor=black!5,linecolor=black!60,linewidth=0.5mm]
    \vspace{0.2mm}
    {\bf Running example: #1.} #2
  \end{mdframed}
}

\begin{document}

\title{\Large \bf \texttt{MixedComplementarityProblems.jl}: A Fast, Batched,
  Open-Source\\Interior Point Solver for Mixed Complementarity Problems}

\author{
  \IEEEauthorblockN{David Fridovich-Keil}
  \IEEEauthorblockA{Department of Aerospace Engineering and Engineering Mechanics\\
  The University of Texas at Austin\\
  Austin, TX, USA\\
  dfk@utexas.edu}
}

\maketitle

\begin{abstract}
Mixed complementarity problems (MCPs) arise as the first-order optimality conditions of
nonlinear programs and noncooperative games, and provide a natural formulation for
multi-agent trajectory optimization problems that appear throughout robotics. The
dominant solver for problems of this form is \texttt{PATH}, which offers strong
performance on robotics problems but remains closed-source. We present \texttt{MixedComplementarityProblems.jl}, an open-source, pure
Julia implementation of an interior point method for parametric MCPs that: (i) matches
\texttt{PATH}'s reliability on standard benchmarks, (ii) natively supports
batched, parallel processing of many parameter instances,
either across CPU threads or on an NVIDIA GPU, and (iii) supports efficient
automatic differentiation of solutions with respect to problem parameters.
On a multi-agent lane-change trajectory game representative of robotics planning
problems, our CPU-multithreaded batched solver clears a batch of parametric instances
 $\sim\!100\times$ faster than sequential calls to \texttt{PATH}. A GPU
backend, running the same solver implementation unmodified, also clears these batches far
faster than \texttt{PATH}, but does not outperform the multithreaded CPU on this problem;
the GPU pulls ahead only once each per-instance KKT system grows large, and we characterize
this regime dependence. We describe the solver's interior point formulation, the
abstraction that lets a single solver implementation run unmodified across dense,
batched-sparse, and single-large linear-algebra backends, and report benchmarks against
\texttt{PATH} on both randomly generated quadratic programs and trajectory games.
\end{abstract}

\section{Introduction}
\label{sec:intro}

Multi-agent trajectory planning problems in robotics, ranging from autonomous driving to human-robot interaction, are naturally posed as
noncooperative dynamic games, in which each agent optimizes its own trajectory subject
to the others' choices. The first-order necessary conditions of optimality for each agent in the game take the form of a
mixed complementarity problem (MCP). Solving MCPs rapidly, therefore, becomes a core computational burden for real-time decision-making in multi-agent robotics applications.

The dominant solver for this class of problems is \texttt{PATH}~\cite{dirkse1995path}. \texttt{PATH}
is mature, reliable, and widely used; but, it is closed-source, solves one problem instance at a
time, and was not designed to support automatic differentiation of solutions with respect to problem parameters.
All three pose serious, practical limitations for use in robotics: (i) developers must have access to solver internals in order to customize to the problem at hand, (ii) scenario- and contingency-based planning (e.g., \cite{li2023scenario,peters2024contingency}) requires solving a \emph{batch} of identically-structured but differently \emph{parameterized} problems, and (iii) modern machine learning pipelines often embed optimization problem solvers as ``layers'' \cite{amos2017optnet}, and end-to-end differentiability requires us to be able to differentiate a solver's output with respect to its input parameters.

\begin{figure}[tp]
  \centering
  \begin{subfigure}{0.48\textwidth}
    \centering
    \includegraphics[width=\linewidth]{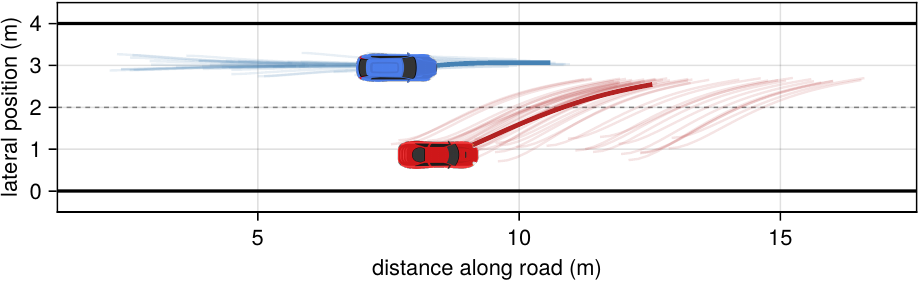}
    \caption{A multi-agent trajectory game is one instance out of a batch of
      parametrically-related MCPs, solved simultaneously. Here, the parameter of
    the game denotes each agent's initial state.}
  \end{subfigure}
  \par\vspace{0.2cm}
  \begin{subfigure}{0.48\textwidth}
    \includegraphics[width=\linewidth]{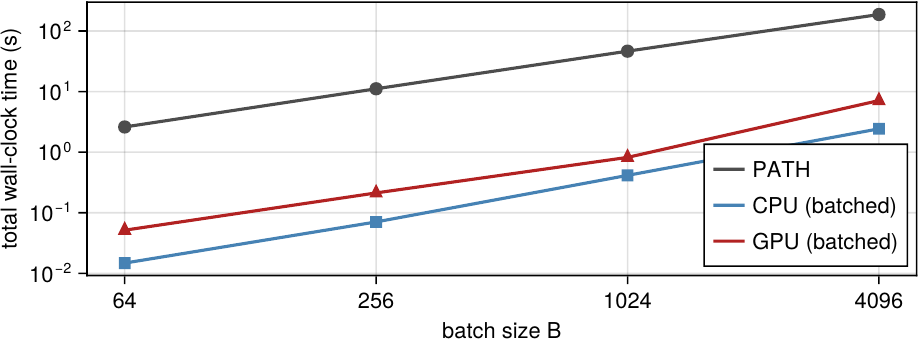}
    \caption{Clearing that batch: sequential \texttt{PATH} calls versus the
      proposed batched CPU/GPU solver.}
  \end{subfigure}
  \caption{Solving a batch of parametric trajectory-game instances (a) with
    \texttt{MixedComplementarityProblems.jl}'s batched interior point solver is
    (b) substantially faster than solving the same batch with sequential calls to
    the existing state of the art, \texttt{PATH} \cite{dirkse1995path}.}
  \label{fig:teaser}
\end{figure}

We present \href{https://github.com/CLeARoboticsLab/MixedComplementarityProblems.jl}{\texttt{MixedComplementarityProblems.jl}}, an open-source, pure Julia solver for parametric MCPs that addresses all of these practical challenges. Concretely, we contribute:
\begin{itemize}
  \item A simple and easily-customized interior point method for parametric MCPs that matches \texttt{PATH}'s
  reliability on standard benchmarks, while supporting automatic differentiation of solutions with respect to problem parameters.
  \item A native \emph{batched} solve mode, in which many parameter instances sharing
  one MCP structure are solved simultaneously, parallelized across CPU threads or an
  NVIDIA GPU.
  \item An abstraction boundary that separates the
    interior point loop and its linear algebra
  backend, which lets a single solver implementation run unmodified across dense,
  batched-sparse, and single-large-instance regimes, so that supporting a new device or
  linear-algebra strategy never requires touching the solver loop itself.
\end{itemize}

As \cref{fig:teaser} previews and \cref{sec:experiments} quantifies, on a
multi-agent lane-change trajectory game representative of robotics planning problems,
clearing a batch of parametric instances with our batched solver is roughly
$100\times$ faster than clearing the same batch with sequential calls to \texttt{PATH}.

\section{Background: Mixed Complementarity Problems}
\label{sec:background}

\subsection{Problem definition}
\label{sec:background:def}

A mixed complementarity problem is specified by a map $F:\mathbb{R}^n\to\mathbb{R}^n$
and bounds $\ell,u\in(\mathbb{R}\cup\{\pm\infty\})^n$; a solution is a vector
$z\in\mathbb{R}^n$ satisfying, for every coordinate dimension $i$,
\begin{equation}
  \begin{aligned}
    \ell_i < z_i < u_i &\implies F_i(z) = 0, \\
    z_i = \ell_i &\implies F_i(z) \ge 0, \\
    z_i = u_i &\implies F_i(z) \le 0.
  \end{aligned}
  \label{eq:mcp-box}
\end{equation}
Readers are referred to \cite{facchinei2003finite} for a complete introduction to MCPs and related problems.

\subsection{MCPs from KKT conditions}
\label{sec:background:kkt}

Consider a parametric nonlinear program
\begin{equation}
  \min_{x} \; f(x;\theta) \quad \text{s.t.} \quad g(x; \theta) = 0,~ h(x;\theta) \ge 0,
  \label{eq:nlp}
\end{equation}
with primal variables $x\in\mathbb{R}^{n_x}$, parameters $\theta$, and Lagrange
multipliers $\lambda \in \mathbb{R}^{n_\lambda}$ associated with the equality constraint $g$ and $\mu \in\mathbb{R}_{\ge 0}^{n_\mu}$ associated with the inequality constraint $h$. The first order necessary (i.e. Karush-Kuhn-Tucker, KKT) conditions of optimality for problem
\cref{eq:nlp} are
\begin{subequations}
  \label{eq:primal-dual-mcp}
  \begin{align}
    &0 = \overbrace{\begin{bmatrix}
    \nabla_x f(x;\theta) - \nabla_x g(x; \theta)^\top \lambda - \nabla_x h(x;\theta)^\top \mu\\
    g(x; \theta)
    \end{bmatrix}}^{G(x,\lambda, \mu;\theta)}  \\
    &0 \le \underbrace{h(x; \theta)}_{H(x;\theta)} \perp \mu \ge 0.
  \end{align}
\end{subequations}

Problem \cref{eq:primal-dual-mcp} can be converted into the form \cref{eq:mcp-box} by taking $z=(x,\lambda,\mu)$, $F = (G, H)$, $u_i = \infty$ ($\forall i$), $\ell_i = -\infty$ ($\forall i \in \{1, 2, \dots, n_x + n_\lambda\}$), and $\ell_i = 0$ ($\forall i > n_x + n_\lambda$).
Because of the natural connection between KKT conditions (e.g., arising from trajectory optimization problems in robotics) and this \emph{primal-dual} form of the MCP, it is the primary form
\texttt{MixedComplementarityProblems.jl} exposes.

\subsection{MCPs from noncooperative games}
\label{sec:background:games}

The construction in \cref{sec:background:kkt} extends from one decision maker to many.
Consider $N$ agents, where agent $i$ solves
\begin{subequations}
\begin{align}
  &\forall i \begin{cases}
  	\min_{x_i} \; f_i(x_i, x_{\neg i};\theta) \\
  	\text{s.t.} \quad g_i(x_i, x_{\neg i}; \theta) = 0,~ h_i(x_i,x_{\neg i};\theta) \ge 0
  \end{cases}\\
  &\text{s.t.} \quad \; g_\mathrm{sh}(x;\theta) = 0,~ h_\mathrm{sh}(x; \theta) \ge 0.
\end{align}
\end{subequations}
Here, each agent $i$ must respect its own private equality and inequality
constraints ($g_i$ and $h_i$, respectively), and also constraints
$g_\mathrm{sh}, h_\mathrm{sh}$ which are shared across agents
(e.g., pairwise collision avoidance).
Each agent $i$'s decision variable is $x_i \in \mathbb{R}^{n_{x_i}}$, and $x_{\neg i}$ denotes the other agents'
decisions so that $x= (x_i, x_{\neg i}) = (x_1,\dots,x_N)$. Writing out each agent's KKT conditions as in
\cref{eq:primal-dual-mcp} and concatenating them---with the shared constraints $g_\mathrm{sh}, h_\mathrm{sh}$
coupling agents through shared multipliers $\lambda_\mathrm{sh}, \mu_\mathrm{sh}$---yields a single primal-dual MCP whose
solution is a generalized Nash equilibrium: a strategy profile from which no agent can
unilaterally improve its own objective \cite{facchinei2003finite}.

Concretely, concatenate the Lagrange multipliers corresponding to equality
constraints as $\lambda \coloneqq
(\lambda_\mathrm{pr}, \lambda_\mathrm{sh})$ and those for inequality
constraints as $\mu \coloneqq (\mu_\mathrm{pr}, \mu_\mathrm{sh})$, with
private multipliers denoted $\lambda_\mathrm{pr} \coloneqq (\lambda_1,\dots,\lambda_N)$ and
$\mu_\mathrm{priv} \coloneqq (\mu_1,\dots,\mu_N)$. Each agent $i$'s Lagrangian is given by 
\begin{multline}
  \label{eq:lagrangian}
  \mathcal{L}_i(x, \lambda, \mu; \theta) \coloneqq f_i(x; \theta) -
  \lambda_i^\top g_i(x; \theta) - \mu_i^\top h_i(x; \theta) - \\
  \lambda_\mathrm{sh}^\top g_\mathrm{sh}(x; \theta) - \mu_\mathrm{sh}^\top h_\mathrm{sh}(x; \theta).
\end{multline}
The $N$ agents' concatenated KKT conditions are then an
instance of \cref{eq:primal-dual-mcp}, with
\begin{equation}
  \begin{aligned}
    G &\coloneqq \left(\nabla_{x_1} \mathcal{L}_1, \dots, \nabla_{x_N} \mathcal{L}_N, g_1,\dots,g_N,
    g_\mathrm{sh}\right), \\
    H &\coloneqq \left(h_1,\dots,h_N,h_\mathrm{sh}\right),
  \end{aligned}
  \label{eq:game-mcp}
\end{equation}
with $G$ and $H$ functions of $(x, \lambda, \mu)$ and parameterized by $\theta$.

\contentbox{A lane-change trajectory game}{
  Consider the two-agent lane changing trajectory game of \cref{fig:teaser}(a).
  Here, each agent's decision variable $x_i = \{s_i(t), a_i(t)\}_{t = 1}^T$ consists of a state-action
  \emph{trajectory} over time horizon $T$. The private equality constraints
  $g_i$ encode system dynamics $\big(s_i(t), a_i(t)\big) \to s_i(t+1)$, private
  inequality constraints encode actuator limits $\|a_i(t)\| \le \bar a_i$, and
  shared inequality constraints encode collision-avoidance $\|P (s_1(t) - s_2(t))\|
  \ge \Delta$ (with projection matrix $P$ recovering agents' position).
  Agents' objective functions $f_i$ encode lane and speed preferences, and
  penalize control effort.
  Parameters $\theta$ encode the agents' initial states $s_i(1)$, lane
  preferences, desired speeds, actuator limits $\bar a_i$, and
  collision-avoidance radius $\Delta$.

  Observe that the shared collision avoidance constraint---and any nonlinearity in agents'
  dynamics---renders the feasible set nonconvex.
}

\section{Solver Design}
\label{sec:design}

\texttt{MixedComplementarityProblems.jl} implements an intentionally-simple
primal-dual interior point method to solve problem \cref{eq:primal-dual-mcp}.
To that end, we introduce a slack variable $w \in \mathbb{R}^{n_\mu}$ and rewrite
\cref{eq:primal-dual-mcp} in terms of $w \ge 0$ and a homotopy parameter $\rho >
0$:
\begin{equation}
\label{eq:slacked-mcp}
K_\rho(x, \lambda, \mu, z; \theta) \coloneqq \begin{bmatrix}
  G(x, \lambda, \mu; \theta) \\
  H(x; \theta) - w \\
  w \odot \mu - \rho 1
\end{bmatrix} = 0.
\end{equation}
Note that the last row of $K_\rho$ involves the elementwise product $w \odot
\mu$ and, consequently, the number of equations in $K_\rho$ is equal to
$d \coloneqq \dim(x) + \dim(\lambda) + \dim(\mu) + \dim(w)$, the total dimension of the
primal, dual, and slack variables---the per-instance KKT system size we return to when
characterizing GPU-vs-CPU scaling in \cref{sec:exp:gpu}.

Per \cref{alg:solver}, we proceed in rounds by solving
\cref{eq:slacked-mcp} via a regularized Newton method and decaying $\rho \to 0$
to satisfy a user-defined tolerance $\|K_0\|
\le \bar K$.
For a given value of $\rho$, the regularized Newton direction $\delta \coloneqq
(\delta_x, \delta_\lambda, \delta_\mu, \delta_w)$ solves
\begin{equation}
  \label{eq:newton}
  \left(\begin{bmatrix}
    \nabla_x G & \nabla_\lambda G & \nabla_\mu G & 0\\
    \nabla_x H & 0 & 0 & -I \\
    0 & 0 & \mathrm{dg}(w) & \mathrm{dg}(\mu)
  \end{bmatrix} + \epsilon I\right) \delta = -K_\rho,
\end{equation}
with $\epsilon \ge 0$ a regularization parameter chosen as described below, and
$\mathrm{dg}(\cdot)$ returning a square matrix with its argument on the main diagonal
and all other entries zero.

We implement two additional subtleties. First, on line~\ref{line:linesearch}, we conduct a
``fraction to the boundary'' linesearch to control the rate at which $w, \mu
\ge 0$ can approach $0$ (cf. \cite[\S19]{nocedal2006numerical}). At each iteration $k$, stepsizes $\alpha_k, \beta_k$ are chosen so that
\begin{equation}
\label{eq:linesearch}
\begin{aligned}
  \alpha_k &= \max\big(\alpha \in [0, 1] ~:~ w + \alpha \delta_w \ge (1 - \tau) w\big)\\
  \beta_k &= \max\big(\beta \in [0, 1] ~:~ \mu + \beta \delta_\mu \ge (1 - \tau) \mu\big),
\end{aligned}
\end{equation}
where the inequalities are interpreted elementwise and $\tau \in (0, 1)$ controls the rate of decay. Linesearch
\cref{eq:linesearch} has a closed-form, exact solution, which makes it amenable
to batched computation as described in \cref{sec:batched}.

Second, we construct an \emph{adaptive} choice of the Newton step regularization
parameter $\epsilon$ and the interior point homotopy parameter $\rho$ that aims
to control subproblem conditioning.
At the end of each inner loop (lines~\ref{line:tighten} and~\ref{line:loosen}), if the solver successfully drove $\|K_\rho\| \le
\rho$ then we \emph{tighten} $\rho$ and $\epsilon$ at a rate dependent upon the
number of Newton steps required to converge; conversely, if the inner loop fails
and $\|K_\rho\| > \rho$, we \emph{loosen} both $\rho$ and
$\epsilon$ to improve the conditioning of the next subproblem and require a less
precise solution.

\begin{algorithm}[t]
\DontPrintSemicolon
\KwIn{initial iterate $(x,\lambda,\mu,w)$ with $\mu,w>0$; parameters $\theta$;
  initial $\rho,\epsilon > 0$; tightening/loosening rates
  $\underline\gamma,\bar\gamma\in(0,1)$; max.\ inner
  iterations $k_\mathrm{max}$; tolerance $\bar K$}
\KwOut{$(x,\lambda,\mu,w)$ with $\|K_0(x,\lambda,\mu,w)\|\le\bar K$}
\While{$\|K_0(x,\lambda,\mu,w;\theta)\| > \bar K$}{
  $k \gets 0$\;
  \Repeat{$\|K_\rho(x,\lambda,\mu,w;\theta)\|\le\rho$ or $k\ge k_\mathrm{max}$}{
    assemble $K_\rho$ \cref{eq:slacked-mcp} and solve \cref{eq:newton} for the Newton direction $\delta=(\delta_x,\delta_\lambda,\delta_\mu,\delta_w)$\;
    compute step sizes $\alpha_k,\beta_k$ via the fraction-to-boundary rule \cref{eq:linesearch}\label{line:linesearch}\;
    $x \gets x+\alpha_k\delta_x$,\quad $\lambda\gets\lambda+\alpha_k\delta_\lambda$,\quad
    $w\gets w+\alpha_k\delta_w$,\quad $\mu\gets\mu+\beta_k\delta_\mu$\;
    $k \gets k+1$\;
  }
  \eIf{$\|K_\rho(x,\lambda,\mu,z;\theta)\|\le\rho$}{
    $\rho \gets \rho\,(1-e^{-\underline\gamma k})$,\quad
    $\epsilon \gets \epsilon\,(1-e^{-\underline\gamma k})$
    \tcp*{tighten}\label{line:tighten}
  }{
    $\rho \gets \rho\,(1+e^{-\bar\gamma k})$,\quad
    $\epsilon \gets \epsilon\,(1+e^{-\bar\gamma k})$
    \tcp*{loosen}\label{line:loosen}
  }
  $\rho \gets \min(\rho, 1)$\;
}
\Return{$(x,\lambda,\mu,z)$}
\caption{Primal-dual interior point method for \cref{eq:primal-dual-mcp}}
\label{alg:solver}
\end{algorithm}

\section{Batched and GPU-Parallel Solving}
\label{sec:batched}

Many applications require solving entire \emph{batches} of MCPs that share
the same symbolic structure $(G, H)$ and differ only in their parameter vector
$\theta$---for example,
\cref{fig:teaser}(a) showcases a batch of trajectory games with varying initial
conditions for the two agents. Rather than naively deploying
\cref{alg:solver} on each instance in series, we reuse the shared symbolic
structure and exploit hardware-level parallelism (both CPU multithreading and
GPU) to solve an entire batch of $B$ parametrically-related instances in a single call.

\contentbox{A batch of lane-change games}{
  Sampling $B$ initial conditions (or lane preferences, etc.)
  as in
  \cref{fig:teaser}(a) yields $B$ instances of the same game, differing only
  in $\theta$. Stacking them column-wise into $\Theta \in \mathbb{R}^{n_\theta \times
    B}$ and calling \cref{alg:batched} solves the entire batch simultaneously,
  e.g., to plan across a distribution of scenarios
  \cite{li2023scenario} rather than one scenario at a time.
}

\subsection{A device- and strategy-agnostic solver loop}
\label{sec:batched:loop}

For convenience, we stack the batch's parameter vectors column-wise into $\Theta \in \mathbb{R}^{n_\theta
\times B}$, so that column $b$ is instance $b$'s parameter vector $\theta_b$;
likewise, we
stack each instance's iterate into $X, \Lambda, M, W \in \mathbb{R}^{(\cdot) \times B}$.
\cref{alg:solver} is then implemented \emph{once}, against a small set of verbs---\texttt{residual!}, \texttt{jacobian!}, \texttt{factorize!}, \texttt{ldiv!}---that
consume and return plain $(\cdot) \times B$ arrays. The numeric representation of the
batched Jacobian $\nabla K_\rho$ and its factorization never leave these four verbs: they
are encapsulated inside a strategy-specific cache, so the solver's control flow is agnostic to
both \emph{where} it runs
and \emph{how} $\nabla K_\rho$ is represented and factored. We implement a
``batched sparse'' strategy: the $B$ instances share one sparsity
pattern for $\nabla K_\rho$, computed once from the symbolic $(G, H)$, and each instance
fills only its own column of the corresponding $(\mathrm{nnz} \times B)$ value
array containing only the structurally nonzero elements of $\nabla K_\rho$.

\subsection{Adapting \cref{alg:solver} to a batch}

Three changes distinguish the batched loop in \cref{alg:batched} from \cref{alg:solver}. First, the homotopy
and regularization parameters $\rho, \epsilon$, and the convergence check
$\|K_\rho\|$, all become length-$B$, one per instance, so that harder
problem instances
can iterate longer without penalizing easier instances in the same batch.

Second, because linesearch \cref{eq:linesearch} has a
closed-form solution, stepsizes are computed elementwise for each instance in the batch. There is no backtracking loop and, in
particular, no per-instance synchronization between host and device.

Third, a batch can often include easy, hard, and infeasible instances.
Therefore, after an instance exceeds a maximum number (e.g., $5$) of consecutive
outer iterations  without satisfying $\|K_\rho\| \le \rho$, it is flagged as a
failure. Depending upon the hardware backend (CPU or GPU), these failures are
either skipped or retained, but kept frozen. This automatic
detection of failures prevents a handful of stalled instances from slowing down
a batch composed of many easier instances.

\begin{algorithm}[t]
\DontPrintSemicolon
\KwIn{batch of $B$ instances $(x_b,\lambda_b,\mu_b,w_b;\theta_b)$ as in
  \cref{alg:solver}, each with its own $\rho_b,\epsilon_b$; tightening/loosening
  rates $\underline\gamma,\bar\gamma$; max.\ inner/outer iterations
  $k_\mathrm{max}, T_\mathrm{max}$; max.\ consecutive stalled outer rounds
  $T_\mathrm{stall}$; tolerance $\bar K$}
\KwOut{status$_b \in \{\texttt{solved}, \texttt{failed}\}$ and
  $(x_b,\lambda_b,\mu_b,w_b)$ for $b = 1,\dots,B$}
status$_b \gets \texttt{active}$, $\mathrm{stall}_b \gets 0$ \quad $\forall b$\;
$t \gets 0$\;
\While{$t < T_\mathrm{max}$ and status$_b = \texttt{active}$ for some $b$}{
  $k \gets 0$\;
  \Repeat{$\|K_{\rho_b}\| \le \rho_b$ for every active $b$, or $k \ge k_\mathrm{max}$}{
    assemble and factor $\nabla K_{\rho_b}$ and solve for $\delta_b$, for every $b$
      with status$_b = \texttt{active}$\label{line:batched-solve}\;
    step active instances via fraction-to-boundary \cref{eq:linesearch}; instances
      with status$_b \ne \texttt{active}$ take a zero step\;
    $k \gets k + 1$\;
  }
  \ForEach{$b$ with status$_b = \texttt{active}$}{
    \eIf{$\|K_{\rho_b}\| \le \rho_b$}{
      \eIf{$\rho_b \le \bar K$}{
        status$_b \gets \texttt{solved}$
      }{
      tighten $\rho_b,\epsilon_b$ (as in \cref{alg:solver}); $\mathrm{stall}_b
      \gets 0$
      }
    }{
      loosen $\rho_b,\epsilon_b$; $\mathrm{stall}_b \gets \mathrm{stall}_b + 1$\;
    }
    \lIf{$\mathrm{stall}_b \ge T_\mathrm{stall}$}{status$_b \gets \texttt{failed}$}
  }
  $t \gets t + 1$\;
}
\Return{$\{\mathrm{status}_b, (x_b,\lambda_b,\mu_b,w_b)\}_{b=1}^B$}
\caption{Batched primal-dual interior point}
\label{alg:batched}
\end{algorithm}

Stalled instances are not left to run: line~\ref{line:batched-solve} restricts
the (re)assembly, factorization, and solve to active instances. This prevents a handful of stalled or failed instances from consuming Newton
iterations once flagged; otherwise, they would force the entire batch to
consume the maximum iteration count.

\subsection{Backends}

On CPU, the shared sparsity pattern is assembled once and each instance's numeric
factorization of $\nabla K_\rho$ reuses the pivot ordering computed on the first
outer round; the $B$ instances are distributed across threads. Because each
instance's factorization is an independent per-instance sparse LU (KLU) \cite{davis2010algorithm}, skipping an
inactive instance at line~\ref{line:batched-solve} is a literal per-instance branch: an
instance with status$_b \ne \texttt{active}$ costs nothing beyond a boolean check, so
CPU throughput scales with the number of instances still active, not $B$.

The GPU backend cannot make the same trade. It factors and solves the batch's shared
sparsity pattern with \texttt{cuDSS}'s \emph{batched} sparse solver, which processes all $B$
instances as a single device-side call and there is no per-instance skip inside a batched
factorization. Consequently, line~\ref{line:batched-solve}'s active-instance
restriction is accepted on GPU only to maintain a uniform verb signature across backends, but
is a no-op there: every outer round factors and solves \emph{all} $B$ instances,
converged and failed ones included, and only their zero stepsize keeps the frozen
instances' $(x_b,\lambda_b,\mu_b,w_b)$ unchanged. This imposes a real cost on
GPU---one paid in exchange for offloading the whole batch to a single, highly
parallel factorization---and it is the reason $T_\mathrm{stall}$ matters more on GPU
than on CPU.

\begin{remark}
The verb set of \cref{sec:batched:loop} is deliberately narrow in order to
readily accommodate other strategies tailored to different problem structures.
For example, a ``batched dense'' strategy could represent
$\nabla K_\rho$ densely, as a $(\cdot) \times (\cdot) \times B$ tensor, and solve it with a batched
LU factorization.
\end{remark}

\section{Differentiable Solving}
\label{sec:differentiable}

Embedding an MCP solve---and the games it can encode---as a differentiable layer inside a
machine learning pipeline requires differentiating the solution with respect to problem
parameters~\cite{amos2017optnet}. Because
\texttt{MixedComplementarityProblems.jl} implements a smooth, interior point method, this capability follows directly from the implicit
function theorem applied to \cref{eq:primal-dual-mcp}, at no cost to the solver design
described above.

Let $z \coloneqq (x, \lambda, \mu, w)$ for the primal-dual-slack iterate. A solution
satisfies $K_0(z; \theta) = 0$ (i.e., \cref{eq:slacked-mcp} in the limit $\rho \to 0$); wherever
$\nabla_z K_0$ is nonsingular, differentiating with respect to $\theta$,
applying the chain rule, and rearranging gives
\begin{equation}
  \label{eq:sensitivity}
  \frac{\partial z}{\partial \theta} = -\left(\nabla_z K_0\right)^{-1} \nabla_\theta K_0.
\end{equation}
Crucially, $\nabla_\zeta K_0$ at a converged iterate is exactly the (unregularized) Newton
Jacobian of \cref{eq:newton} already assembled and factored at the last step of
\cref{alg:solver}. Evaluating \cref{eq:sensitivity} therefore costs one additional
$n_\theta$-column \texttt{ldiv!} against that same factorization---not a new factorization---and
extends unchanged to the batched setting, where it becomes a per-instance multi-right-hand-side
solve of shape $(\cdot) \times n_\theta \times B$ against each instance's existing factorization.

Equation \cref{eq:sensitivity} supports both forward- and reverse-mode automatic differentiation. Forward-mode
autodifferentiation dispatches on dual number-typed $\theta$: the solver solves once at the
underlying real value and propagates dual perturbations through $\nicefrac{\partial z}{\partial
\theta}$. Reverse-mode autodifferentiation is implemented via a single
Jacobian-vector product. Both are implemented for the unbatched and batched
solvers alike, and because
$\nabla_\theta K_0$ is only needed for \cref{eq:sensitivity}, its symbolic construction is
an opt-in user choice during MCP construction, so users who do not need
gradients pay no additional compile- or run-time cost.

As with any KKT-based sensitivity, \cref{eq:sensitivity} presumes the set of
active constraints for each problem is constant near $\theta$: where an
inequality constraint is only weakly active, the true solution map
is only Lipschitz, not differentiable, and $\nicefrac{\partial z}{\partial \theta}$ is a one-sided
approximation. This is a property of the KKT system itself, not a solver-specific limitation.

\contentbox{Differentiating the lane-change game}{
  We validate \cref{eq:sensitivity} against central finite differences on the lane-change
  game of \cref{fig:teaser}(a). Over $20$ random instances converging to $\|K_0\| \le
  10^{-4}$, differentiating a scalar function of the solution (the primal and dual
  variables' squared norm) with respect to $\theta$ via forward-mode
  autodifferentiation (using \texttt{ForwardDiff.jl} \cite{RevelsLubinPapamarkou2016}) agrees with a
  central-difference approximation (step $10^{-6}$) to a median relative error, in the
  gradient's norm, of $6 \times 10^{-9}$, and no worse than $0.1\%$ across all $20$
  instances.
}

\section{Experiments}
\label{sec:experiments}

We evaluate \texttt{MixedComplementarityProblems.jl} against \texttt{PATH} along the two
axes that matter for the robotics applications of \cref{sec:intro}: reliability---\emph{does the solver converge as often as \texttt{PATH}?}---and throughput---\emph{how quickly
can it clear a batch of parametrically-related instances?} Concretely, our experiments
answer four questions:
\begin{enumerate}
  \item \textbf{Single-instance parity.} Solving one problem at a time, does our
  interior point method match \texttt{PATH}'s reliability, and how does its per-solve
  wall-clock timing compare? (\cref{sec:exp:single})
  \item \textbf{Batched throughput.} When a whole batch must be cleared, how much faster
  is a single batched solve than sequential \texttt{PATH} calls? (\cref{sec:exp:throughput})
  \item \textbf{CPU thread scaling.} How does batched throughput scale with the number of
  CPU threads? (\cref{sec:exp:threads})
  \item \textbf{GPU vs.\ CPU.} When does offloading the batch to a GPU beat a
  many-threaded CPU, and why? (\cref{sec:exp:gpu})
\end{enumerate}

\subsection{Experimental setup}
\label{sec:exp:setup}

\paragraph{Problem families}
We benchmark on two families of parametric MCPs. The first is a family of randomly
generated sparse, convex \emph{quadratic programs} (QPs) of the form $\min_x \tfrac12
x^\top M x - \phi^\top x$ s.t.\ $Ax - b \ge 0$, with $M \succeq 0$; the parameter vector
$\theta = (M, A, b, \phi)$ is drawn at random with a controllable sparsity rate, and the
problem dimensions (numbers of primal variables $n_x$ and inequality constraints
$n_\mu$) are free hyperparameters. This family exercises the solver on a controlled range of KKT
system sizes and, because the random instances are not guaranteed to be feasible, on a realistic
mix of solvable and infeasible instances. The second family is the lane-change
\emph{trajectory game} of \cref{sec:background:games} (the running example), a two-player
generalized Nash equilibrium problem over a planning horizon $T \in \mathbb{N}$; each sampled instance
randomizes the players' initial states, so the batch spans a
distribution of scenarios exactly as a scenario-based planner would encounter
\cite{li2023scenario}. The horizon $T$ controls the per-instance KKT dimension.

\paragraph{Baselines}
The primary baseline is \texttt{PATH}~\cite{dirkse1995path}, accessed through
\texttt{PATHSolver.jl} \cite{PATHSolver} and \texttt{ParametricMCPs.jl} \cite{ParametricMCPs}; \texttt{PATH} solves one instance
at a time on a single thread. We additionally report our own \emph{unbatched}
interior point solver (\cref{alg:solver}) as an intermediate baseline, isolating
the performance of the interior point formulation from the batching machinery. Against these, we
compare the batched interior point solver (\cref{alg:batched}) in both its CPU
(multithreaded) and GPU (\texttt{cuDSS}) configurations.

\paragraph{Hardware and software}
All experiments use Julia 1.12.6 and run on a single workstation with an AMD Ryzen 9 7950X CPU (16 physical
cores / 32 hardware threads) and 126~GB of RAM; GPU experiments use an NVIDIA GeForce
RTX~4090 (24~GB). All timings are taken after a
separate warm-up solve to exclude compilation. Per-solve times (cf.  \cref{tab:single}) are
summarized as mean~$\pm$~standard deviation over the $B$ instances of a batch; batched
throughput (cf. \cref{tab:throughput}) is the total wall-clock to clear the batch, which varies
modestly across repetitions (standard deviation under about $20\%$). Unless noted, we use a
convergence tolerance of $\bar K = 10^{-4}$ and report each batch's \emph{solved fraction}
(instances reaching $\|K_0\| \le \bar K$) and its total wall-clock time to clear all $B$
instances. For a fair comparison, \texttt{PATH} is held to the same convergence tolerance.

\subsection{Single-instance reliability and speed}
\label{sec:exp:single}

Before batching, we evaluate \cref{alg:solver} against \texttt{PATH} on individual
(unbatched) problems. \cref{tab:single} reports, for each problem family, the fraction of
instances each solver converges on and its per-solve wall-clock timing, over $1024$ random
instances (QP: $n_x = 32$, $n_\mu = 16$; trajectory game: $T = 10$).
Two results stand out. First, \cref{alg:solver} is as reliable as \texttt{PATH}:
it converges on $431$ of the $1024$ QP instances to \texttt{PATH}'s $437$ (the remainder
being infeasible), and on \emph{all} $1024$ trajectory games, where \texttt{PATH} declares $28$ failures.
Second, in contrast to what one might expect against a mature, heavily optimized
solver, once its Newton system is factored with \texttt{KLU} (see below), \cref{alg:solver}
is per-solve \emph{competitive with, and often faster than}, \texttt{PATH}: its median solve
is roughly $7\times$ faster on the QP and $55\times$ faster on the trajectory game. On the
QP its \emph{mean} is nonetheless comparable to \texttt{PATH}'s: roughly half of the randomly
generated instances are infeasible, and \cref{alg:solver} iterates each to its round limit
before declaring failure, a heavy tail that inflates its own mean well above its median
(hence its large standard deviation in \cref{tab:single}). \texttt{PATH}'s solve times are
similarly bimodal, though its particular mean--median relationship is sensitive to exactly
how and when it detects infeasibility.

The sparse linear solver used inside \cref{alg:solver} matters substantially here.
\cref{alg:solver} refactorizes the regularized Jacobian of \cref{eq:newton} at every Newton
step, always with the same sparsity pattern. \texttt{KLU}
\cite{davis2010algorithm} and \texttt{UMFPACK} \cite{davis2004algorithm} are both
standard direct solvers for sparse linear systems; \texttt{KLU}'s in-place refactorization,
which reuses the symbolic analysis and pivot ordering across steps, is $4$--$6\times$ faster
per solve than \texttt{UMFPACK} on these repeatedly refactored systems, at identical
reliability (cf. \cref{tab:single}). We therefore adopt \texttt{KLU} by default; it is also the
factorization the batched CPU backend uses (cf. \cref{sec:batched}).

\begin{table}[t]
  \centering
  \begin{tabular}{llcc}
    \toprule
    Problem & Solver & Solved & Time (ms) \\
    \midrule
    \multirow{3}{4em}{Random QP}
      & \texttt{PATH}               & $437/1024$ & $3.9 \pm 3.6$ \\
      & \cref{alg:solver} + \texttt{UMFPACK} & $431/1024$ & $17.7 \pm 19.6$ \\
      & \cref{alg:solver} + \texttt{KLU}     & $431/1024$ & $4.7 \pm 6.3$ \\
    \midrule
    \multirow{3}{4em}{Trajectory game}
      & \texttt{PATH}               & $996/1024$ & $45.2 \pm 22.3$ \\
      & \cref{alg:solver} + \texttt{UMFPACK} & $1024/1024$ & $5.5 \pm 4.4$ \\
      & \cref{alg:solver} + \texttt{KLU}     & $1024/1024$ & $0.9 \pm 0.5$ \\
    \bottomrule
  \end{tabular}
  \caption{Single-instance reliability and speed: solved fraction and per-solve wall-clock
    (mean~$\pm$~std) over $1024$ random instances of each family (QP: $n_x = 32$,
    $n_\mu = 16$; trajectory game: $T = 10$). The infeasible QP instances make the
    per-solve distribution bimodal, so for \cref{alg:solver} the mean sits well above the median (QP medians:
    \texttt{PATH} $4.1$, \cref{alg:solver} + \texttt{UMFPACK} $2.5$,
    \cref{alg:solver} + \texttt{KLU} $0.6$~ms).}
  \label{tab:single}
\end{table}

\subsection{Batched CPU throughput vs.\ \texttt{PATH}}
\label{sec:exp:throughput}

The batched solver's purpose is to clear an entire batch at once. \cref{tab:throughput}
reports the total wall-clock to solve a batch of $B = 1024$ instances of each family
three ways: sequential \texttt{PATH} calls, sequential unbatched \cref{alg:solver}
calls, and via a single \cref{alg:batched} call across $32$ CPU threads.
As previewed in \cref{fig:teaser}(b), on the trajectory game the CPU-multithreaded batched
solver clears the batch roughly $100\times$ faster than sequential \texttt{PATH}; even the
\emph{unbatched} solver is already $\sim\!50\times$ faster than \texttt{PATH} here, because
its per-solve cost is so much lower (cf. \cref{tab:single}), and batching then adds a further
$\sim\!2\times$ on top.

On the QP the batched solver is $6.8\times$ faster than \texttt{PATH}, even though the
\emph{sequential} unbatched solver is here slightly slower than \texttt{PATH} in total
(cf. \cref{tab:throughput}). This is a consequence of the infeasible instances. Roughly half of
the random QPs are infeasible, and each such instance runs all the way to
\cref{alg:solver}'s outer-iteration limit before it is declared a failure; these few
expensive solves dominate the \emph{sequential} total, even though the typical (feasible)
solve is fast, per \cref{tab:single}. The batched solver absorbs this tail, with the per-instance stall
detection of \cref{alg:batched} freezing them early so that they no longer gate
the total runtime on CPU. The
batched solver therefore clears the whole batch $6.8\times$ faster than \texttt{PATH}.

The batched speedup comes from two compounding sources: sharing the symbolic
$(G, H)$ structure across the batch (assembled and sparsity-analyzed once), and
parallelizing the per-instance factorize/solve across threads. Comparing against the
sequential \cref{alg:solver} isolates the second effect---due purely to
parallelization---from the first: it is $9.7\times$ faster on the QP but only
$2.2\times$ faster on the
game, where each per-instance solve is already inexpensive, so there is less serial work
for threading to absorb.

\begin{table}[t]
  \centering
  \begin{tabular}{llccc}
    \toprule
    Problem & Solver & Total time & Solved & Speedup \\
    \midrule
    \multirow{3}{*}{Random QP}
      & \texttt{PATH}          & $3.97$~s & $437/1024$ & $1\times$ \\
      & \cref{alg:solver}      & $5.61$~s & $431/1024$ & $0.7\times$ \\
      & \cref{alg:batched}     & $0.58$~s & $424/1024$ & $\mathbf{6.8\times}$ \\
    \midrule
    \multirow{3}{*}{Trajectory game}
      & \texttt{PATH}          & $46.4$~s & $996/1024$ & $1\times$ \\
      & \cref{alg:solver}      & $0.97$~s & $1024/1024$ & $47.8\times$ \\
      & \cref{alg:batched}     & $0.44$~s & $1024/1024$ & $\mathbf{105.5\times}$ \\
    \bottomrule
  \end{tabular}
  \caption{Batched throughput on CPU. Total wall-clock time and solved fraction for clearing a batch
    of $B = 1024$ instances, and speedup over sequential \texttt{PATH}. Both our solvers use
    the \texttt{KLU} factorization in solving \cref{eq:newton}.}
  \label{tab:throughput}
\end{table}

\begin{figure}[bp!]
  \centering
  \includegraphics[width=0.9\columnwidth]{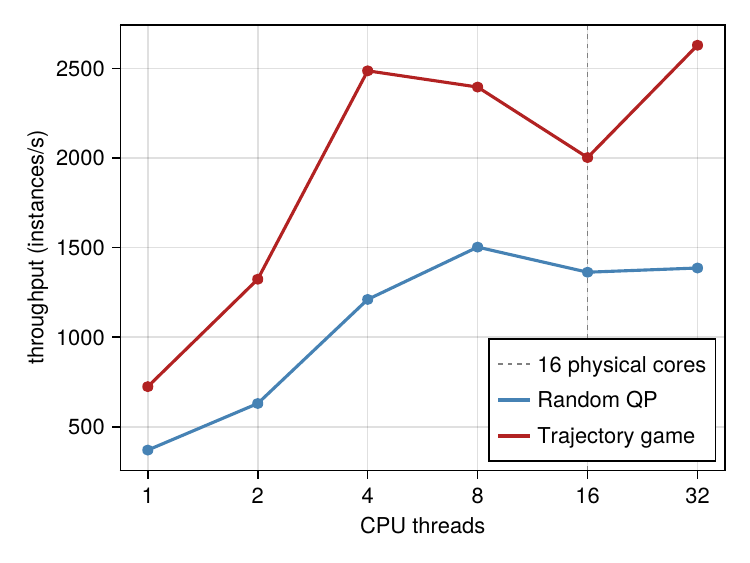}
  \caption{Batched-solver throughput vs.\ CPU thread count at a fixed batch size of $N = 1024$.
    Throughput rises sublinearly with the thread count, with the gains concentrated in the first
    few threads; both problem families plateau around physical-core saturation, with no further
    systematic gain from simultaneous multithreading.}
  \label{fig:threadscaling}
\end{figure}

\begin{figure*}[tp!]
  \centering
  \includegraphics[width=0.85\textwidth]{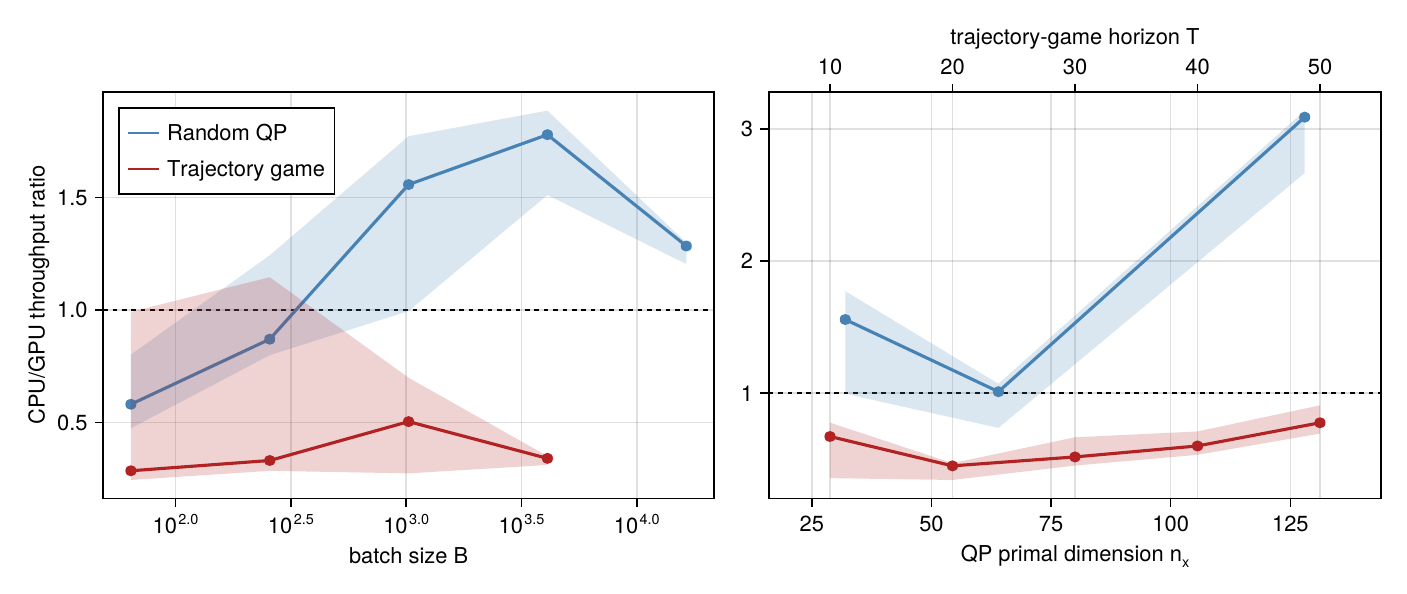}
  \caption{GPU (\texttt{cuDSS}) vs.\ CPU batched throughput. The GPU pulls ahead only for
    large, dense per-instance systems (the QP at large $n_x$ or large $B$); on the sparse
    trajectory-game systems the multithreaded CPU is faster at every horizon, because its
    active-set skip factorizes only the still-active instances each step while \texttt{cuDSS}
    always processes the whole batch. Shaded bands show a conservative envelope,
    $[\min(\text{CPU})/\max(\text{GPU}), \max(\text{CPU})/\min(\text{GPU})]$, over $5$--$30$
    repeated trials per point.}
  \label{fig:gpu}
\end{figure*}

\subsection{CPU thread scaling}
\label{sec:exp:threads}

\cref{fig:threadscaling} shows batched throughput (instances cleared per second) as a
function of the number of CPU threads, at a fixed batch size of $N = 1024$ for both problem
families. Because each instance's sparse factorization is independent (cf. \cref{sec:batched}),
throughput increases with the thread count, though sublinearly. The gains are concentrated in
the first few threads: on the QP, throughput rises from $371$ instances/s single-threaded to
$1211$ at four threads and peaks around $1500$ instances/s ($4\times$), while the trajectory
game rises from $725$ to $2486$ instances/s at four threads ($3.4\times$) and then holds near
that level. Beyond roughly the first four-to-eight threads both curves flatten: adding threads
up to and past the $16$ physical cores yields no further systematic gain, only run-to-run
noise, as simultaneous multithreading does not add useful parallelism for this factorization-bound
workload. The game sustains higher absolute throughput than the QP despite the smaller
speedup, because each game instance converges in fewer Newton iterations, whereas the QP batch
is roughly $60\%$ infeasible-by-construction instances that are more expensive to rule out. In
both cases the batched solver extracts substantial parallel throughput from commodity multicore
hardware with no device-specific code.

\subsection{GPU vs.\ CPU: a regime-dependent crossover}
\label{sec:exp:gpu}

Whether the GPU backend beats the multithreaded CPU depends on the problem regime.
Because both backends run the same solver loop behind the verb set of \cref{sec:batched},
this comparison isolates the linear-algebra backend. We sweep two axes independently
in \cref{fig:gpu}: the batch size $B$ at fixed per-instance size (left), and the
per-instance size at fixed $B$ (right)---controlled by the number of primal variables
$n_x$ in the QP and by the horizon $T$ in the trajectory game.

GPU outperforms batched CPU only once each per-instance KKT system is both large
\emph{and} sufficiently dense.
On the QP, the GPU backend pulls ahead as the batch grows larger (up to $1.8\times$ faster than the $32$-thread
CPU at $B = 4096$) and, more strongly, as the per-instance system grows: at $n_x = 128$ the
GPU is roughly $3\times$ faster. On the trajectory game, by contrast, the CPU is faster at
\emph{every} planning horizon we tested (median of five repetitions): at $T = 30$
($d \approx 2000$) the GPU takes $8.0$\,s to the CPU's $4.1$\,s ($1.9\times$ slower), and at
$T = 50$ ($d \approx 3500$), $16.3$\,s to $12.7$\,s. Both backends nonetheless clear these
batches far faster than sequential \texttt{PATH}.

On its face, these results appear counterintuitive: the trajectory game KKT systems are the \emph{largest} we solve,
yet here the GPU backend is uniformly slower than batched CPU. Two opposing effects govern
end-to-end cost, and they separate cleanly when the batched linear-algebra verbs
(\texttt{jacobian!}, \texttt{factorize!}, \texttt{ldiv!}) are timed in isolation.
\emph{Per Newton step, with the whole batch active}, the GPU's batched sparse factorization
does become cheaper than the CPU's per-thread \texttt{KLU} as each instance's Jacobian grows:
its combined assemble-and-factorize cost crosses over the CPU's at $d \approx 2500$
($T \approx 35$) and is $\sim\!1.4\times$ cheaper by $d \approx 4900$, because \texttt{cuDSS}
amortizes its fixed launch and occupancy overhead better as the matrix grows while the CPU
factorization has no comparable fixed cost. \emph{However, a full solve does not run with the whole
batch active.} \cref{alg:batched}'s set of active instances shrinks as instances converge or are frozen,
and on the CPU each step factorizes only the still-active instances (\cref{sec:batched}),
whereas on GPU \texttt{cuDSS} always processes the entire batch. Measured at $d = 3500$, the GPU's
per-step advantage of $1.3\times$ with all $1024$ instances active inverts to an $8\times$
\emph{disadvantage} once only $32$ remain active. Since a real solve spends most of its steps
with a small group of active instances, the CPU wins end-to-end.
The GPU therefore prevails only when each per-instance system is large enough for
\texttt{cuDSS} to win the per-step comparison \emph{and} dense enough---as in the large-$n_x$
QP---that its factorization advantage is large; the trajectory game KKT
systems meet the first condition but not the second.

Two further caveats bear mention. First, the fraction of solved instances in a
batch of trajectory games declines at longer
time horizons (for our specific test problem), even with warm starting ($92\%$ at $T = 30$, $57\%$ at $T = 50$), and cold starting
collapses it further, so we restrict the game comparison to the warm-started
regime and caution against over-interpretation of results on very long horizons.
Second, the fraction of solved instances in a batch of QPs changes sharply with $n_x$ ($41\% \to 96\% \to 100\%$ from
$32$ to $128$ primals), which complicates the interpretation of those timing results.

\section{Discussion and Future Work}
\label{sec:discussion}

Taken together, our experiments support a simple practical message: for the parametric MCPs
that arise in multi-agent robotics, the decisive speedup comes from \emph{batching} the solve,
not from any single choice of hardware. A CPU-multithreaded batch clears a set of trajectory-game
instances roughly two orders of magnitude faster than sequential calls to \texttt{PATH}
(\cref{sec:exp:throughput}) while matching its per-instance reliability
(\cref{sec:exp:single})---and it does so on commodity hardware, with no device-specific code.
The verb-set abstraction of \cref{sec:batched} is what makes this portable: the same interior
point loop runs unmodified across both CPU and GPU backends.

That abstraction also lets us characterize \emph{when} each device is the right choice, and the
answer is regime-dependent (cf. \cref{sec:exp:gpu}). The multithreaded CPU is the better default for
the extremely sparse KKT systems typical of receding-horizon trajectory games: its
active instance logic factorizes only the still-active instances each Newton step, whereas \texttt{cuDSS}
always processes the whole batch. The GPU outperforms only once each per-instance system is both
large and dense enough for its batched factorization to win the per-step
comparison outright---e.g., the
large-$n_x$ QP. Future work should close this gap by
making the GPU-batched factorization aware of the set of active instances, e.g.\ compacting the
active instances before each \texttt{cuDSS} call so that GPU per-step cost, like the CPU's, falls
as the number of active instances shrinks; cf. \cite{viljoen2026scaling}, which
implements a similar design (but for nonlinear programs rather than mixed
complementarity problems).

Several other limitations temper these conclusions and motivate further work. Our reliability and timing
comparisons use two problem families---randomly generated QPs and a single lane-change trajectory
game---and, on the trajectory game, the fraction of solved instances declines at long horizons even with warm
starting (cf. \cref{sec:exp:gpu}). A broader benchmark suite would improve the
prescriptive quality of our evaluations, and a more
sophisticated interior point implementation is needed to improve robustness. Quantifying the
wall-clock cost of differentiating through a batched solve (\cref{sec:differentiable}) at the
scale required by large learned pipelines, and integrating it end-to-end into such a
pipeline~\cite{amos2017optnet}, remains important future work. Finally,
integrating the batched solver into a deployed receding-horizon planner would test the
warm-started, real-time regime that these throughput numbers are meant to serve.

\bibliographystyle{IEEEtran}
\bibliography{refs}

\end{document}